\documentstyle [twocolumn,epsf]{mn}
\newcommand{\mincir}{\raise
-3.truept\hbox{\rlap{\hbox{$\sim$}}\raise4.truept\hbox{$<$}\ }}
\newcommand{\magcir}{\raise
-3.truept\hbox{\rlap{\hbox{$\sim$}}\raise4.truept\hbox{$>$}\ }}
\newcommand{\minmag}{\raise
-3.truept\hbox{\rlap{\hbox{$<$}}\raise5.truept\hbox{$<$}\ }}
\newcommand{\be}{\begin{equation}}
\newcommand{\ee}{\end{equation}}
 \newcommand{\ba}{\begin{eqnarray}}
\newcommand{\ea}{\end{eqnarray}}
\newcommand{\brr}{\begin{array}}
\newcommand{\err}{\end{array}}
\newcommand{\bc}{\begin{center}}
\newcommand{\ec}{\end{center}}

\newcommand{\Msun}{\,\mbox{${\rm M}_\odot$}}

\title[The Shape-Alignment relation in $\Lambda$CDM Cosmic
Structures]
{The Shape-Alignment relation in $\Lambda$CDM Cosmic Structures}
\author[Basilakos et al.]{S. Basilakos$^{1,2}$, M. Plionis$^{1,3}$, 
G. Yepes$^4$, S. Gottl\"{o}ber$^5$, V. Turchaninov$^6$.\\
\vspace{0.1cm}
$^1$ Institute of Astronomy \& Astrophysics, National Observatory of Athens, 
I. Metaxa \& V. Pavlou, Palaia Penteli, 15236 Athens, Greece \\
$^2$ Kapteyn Astronomical Institute, University of Groningen, The Netherlands\\
$^3$ Instituto Nacional de Astrofisica, Optica y Electronica (INAOE)
Apartado Postal 51 y 216, 72000, Puebla, Pue., Mexico \\
$^4$ Grupo de Astrof\'\i sica, Universidad Aut\'onoma de Madrid, 
Madrid E-28049, Spain \\
$^5$ Astrophysikalisches Institut Potsdam, An der Sternwarte 16,
14482 Potsdam, Germany \\
$^6$ Keldysh Institute for Applied Mathematics,
Miusskaja Ploscad 4,  125047 Moscow, Russia
}

\begin{document}

\maketitle

\begin{abstract}
In this paper we study the supercluster - cluster morphological 
properties using one of the largest ($2\times 512^{3}$ 
SPH+N-body simulations of large scale structure formation
in a $\Lambda$CDM model, based
on the publicly available code GADGET.
We find that filamentary (prolate-like) shapes are the 
dominant supercluster and cluster dark matter halo 
morphological feature, in agreement with 
previous studies. However, the baryonic gas component of the clusters is
predominantly spherical.
We investigate the alignment between
cluster halos (using either their DM or baryonic components) 
and their parent supercluster major-axis orientation, finding that
clusters show such a preferential alignment.
Combining the shape and the alignment statistics, we also 
find that the amplitude of supercluster - cluster alignment increases
although weakly with supercluster filamentariness.
\end{abstract}

\begin{keywords}
galaxies: clusters: general - 
cosmology:theory - large-scale structure of universe 
\end{keywords}

\vspace{1.0cm}

\section{Introduction}
The wealth of recent observational data
has brought great progress in understanding the
cosmic structure formation pattern. At the largest cosmic scales
it is known that clusters of galaxies are not randomly 
distributed but tend to aggregate in even larger conglomerations, 
the so called superclusters (cf. Oort 1983; Bahcall 1988).
The large sizes of superclusters ($\magcir 30 \; h^{-1}$ Mpc)
in conjunction with the amplitude of their member cluster peculiar
velocities ($\sim$ 1000 km/sec) imply that superclusters
constitute unvirialized density fluctuations which should reflect
the initial conditions that gave rise to structure formation
processes.

Observational and theoretical studies of the supercluster 
morphology and environment suggest that 
they are not spherical but instead their shapes are elongated 
(eg. Zeldovich, Einasto \& Shandarin 1982; 
de Lapparent, Geller \& Huchra
1991) with a prolate-like tendency (eg. Frenk et al. 1988; 
Plionis, Valdarnini \& Jing 1992; Dubinski 1994;
Einasto et al. 2001; Jing \&  Suto 2002;
Diaferio et al. 2003; 
Sheth et al. 2003; Einasto et al. 2003; 
Shandarin, Sheth, Sahni 2004; 
Pandey \& Bharadwaj 2005). 
Sathyaprakash, Sahni \& Shandarin (1998), 
Basilakos, Plionis \& Rowan-Robinson
(2001), Kolokotronis, Basilakos \& Plionis (2002), using either the
IRAS galaxy distribution or the Abell/ACO clusters, gave attention to 
the cosmological inferences of supercluster shape statistics claiming that a 
low matter density flat cosmological model ($\Omega_{\rm
m}=1-\Omega_{\Lambda}=0.3$) fits the large-scale 
observational results at a high significance level. 

A variety of indications support the formation of
clusters by hierarchical aggregation of smaller units along filamentary
large-scale structures (eg. West 1994; Ostriker \& Cen 1996 and 
references therein). Under this scenario dynamically young
clusters will have a tendency to be aligned with neighboring
structures, since the accretion of matter takes place along the
large-scale filamentary structure within which the clusters form.
Indeed, observational studies of structure orientations, which 
has a long history in cosmology, have found strong indications of
various alignment effects.
Bingelli (1982) was the first to find that the major axes of neighboring
clusters of galaxies tend to point toward each other. 
On the other hand, Struble \& Peebles (1985) have
failed to measure a significant alignment signal (see also 
Flin 1987; Rhee \& Katgert 1987; Ulmer, McMillan, \& Kowalski 
1989). However, in the last decade many authors have claimed that 
cluster formation processes are
strongly connected to the supercluster network and thus 
generate measurable environmental effects, among
which strong alignment effects observed in both observational and
N-body data (eg. West 1994; Plionis 1994; West, Jones, \& Forman 1995;  
van Haarlem, Frenk \& White 1997;  Colberg et al. 2000; 
Onuora \& Thomas 2000; 
Chambers, Mellot \& Miller 2002; Plionis \& Basilakos 2002; 
Faltenbacher et al. 2002; Plionis et al. 2003; Bailin \& Steinmetz 2004; Kasun
\& Evrard 2004; Pimbblet 2005; Hopkins, Bahcall \& Bode 2005; Lee,
Kang \& Jing 2005; Faltenbacher et al. 2005). 

Furthermore, Plionis (2002; 2004) using the APM galaxy data,
has shown that the alignment effects are not
confined only between cluster pairs or between luminous galaxies and
their parent clusters, but extend to alignments between clusters and
their parent supercluster major axis. This again is the sort of picture that
one may expect in hierarchical structure formation scenarios where
matter and galaxies flow within one or two dimensional structures, 
on the intersects of which the clusters form.

The aim of this work is along the same lines, attempting 
to make a detailed investigation of the connection between the 
morphological and environmental properties of the large scale structure in
the concordance $\Lambda$CDM cosmology. For a detailed study of the
morphological properties of the supercluster-void network in such a
cosmological model we refer the reader to the recent work of Shandarin, Sheth
\& Sahni (2004).

The plan of the paper is as follows. The simulated cluster samples are 
presented in section 2.  In section 3, we briefly describe the 
method used to investigate supercluster and cluster shape 
properties and comment on some systematic effects related to the
definition of superclusters. In section 4, we discuss the
large scale structure orientation effects and finally
in section 5, we present our conclusions.

\section{The simulated $\Lambda$CDM clusters}
In this study we use large-scale N-body simulations 
of the popular $\Lambda$CDM  cosmological model in order to quantify 
the 3D cluster and supercluster morphological properties and their
relation to the large scale structure network.

\subsection{N-body Simulation}
\label{nbody}

We have performed numerical simulations of a 500 $h^{-1}$ Mpc cubic volume
in which a random realisation of the concordance $\Lambda$CDM
$(\Omega_{\rm m}=0.3, \Omega_{\Lambda} = 0.7, h=0.7, \Omega_B=0.045,
\sigma_8=0.9)$ power spectra was generated with $2048^3$ particles.
Due to computational limitations, we have not yet simulated down to these
resolutions but only selected regions. For the whole box, 
we increased the mass resolution in steps of a
factor of 8 by averaging 
$2048^3$ initial distribution of particles. Then we replace each
particle
by a dark matter and gas particle. We have run simulations with 
$2\times 256^3$ and $512^3$ particles.
Since all simulations were extracted from the
same initial conditions, the same structures are formed, so that we can
estimate the effects of mass resolution on the results (see \S \ref{resol}).

The simulations were done with an updated version of the parallel
Tree-SPH code
GADGET (Springel, Yoshida \& White 2000). The code uses
an entropy-conserving formulation of SPH (Springel \& Hernquist 2002)
which alleviates problems due to numerical overcooling.
In order to study the gas
dynamics of clusters, we ran the simulations with the same number of
sph and dark matter particles. The results reported in this paper are
based on the highest resolution simulation carried out which consists
of $2 \times 512^3 (\sim 10^{8.43}) $ particles with a mass resolution
of $m_{\rm dark} = 6.6 \times 10^{10} h^{-1} \Msun $ and $m_{\rm sph} = 1.2
\times 10^{10} h^{-1} \Msun $.  We still have a factor of 64 in mass
left until we reach the resolution limit of the initial conditions.
This allows further improvement in numerical results on larger
computers. Anyway, our simulation is already one of the largest
adiabatic SPH simulations of large scale structure done so far.  The
mass resolution of the $2 \times 512^3$ simulation allows us to
reliably identify from large galactic halos (100+100 particles) to the
biggest galaxy clusters ($4 \times 10^4$ particles).

The spatial force resolution was set to an equivalent Plummer gravitational
softening of $15 \;h^{-1}$ comoving kpc.  
The sph smoothing length was set to the  distance to the 40$^{th}$
nearest neighbour of each sph particle. In any case, we do not allow
smoothing scales to be smaller than the gravitational softening of the gas
particles. 

These experiments have also been used in order to re-simulate individual
clusters found in the low resolution run with full resolution to study
the properties of the ICM in hot X-ray clusters (Yepes et al. 2004).
The number density of
clusters found in this simulation agree quite well with recent
observations of the X-ray temperature function by Ikebe et al. (2002)
ranging from 1 keV to 10 keV clusters (for further details see Yepes
et al. 2004). 

The simulations were performed in  the IBM Regatta p690+ supercomputer
cluster  at the John von Neumann Center J\"ulich. For the
largest simulation of $2\times 512^3$ particles, we employed  64 cpus
simultaneously  for a total  of $64\times 17=10000$ CPU hours.

\subsection{Simulated Cluster Samples}
At first we have calculated the minimal spanning tree (MST) of the DM
particle distribution. The minimal spanning tree of any point
distribution is a {\em unique}, well defined quantity which describes
the clustering properties of the point process completely (e.g.,
Bhavsar \& Splinter 1996, and references therein). The minimal spanning
tree of $N$ points contains $N-1$ connections. We are using an MPI
program which calculates the MST of the $512^3$ particles using 8 CPUs
within about 10 minutes. In a second step we cut the MST using
different linking lengths in order to extract catalogs of
friends-of-friends particle clusters. Note, that cutting a given MST is a
very fast algorithm.  We start with a linking length of 0.17 times the
mean inter particle distance which corresponds roughly to objects with
the virialization overdensity $\rho/\rho_{\rm mean}\simeq 330$ at
($z=0$).  Decreasing the linking length by a factor of $2^n$ ($n =
$1,2,...) we get samples of objects with roughly $8^n$ times larger
overdensities which correspond to the inner part of the objects of the
first sample. With this hierarchical friends-of-friends algorithm we
can also detect substructures of clusters (cf. Klypin et al. 1999).

With a linking length of 0.17 we find $\sim77000$ objects in the
simulation box with more than 100 dark matter particles , i.e. total 
masses $m_{\rm fof} \geq 8\times 10^{12} h^{-1} \Msun $.  The
corresponding mass function is shown in Fig. 1
together with the Sheth and Tormen (1999) analytical approximation.

\begin{figure}
\mbox{\epsfxsize=8.cm \epsffile{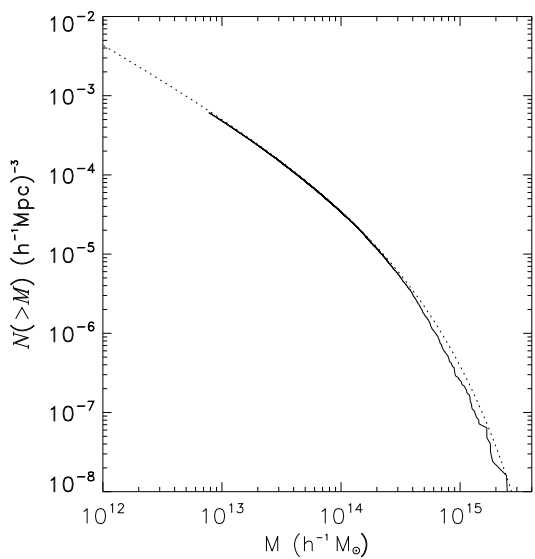}}
\caption{The mass function of FOF objects in the $500  h^{-1}$ Mpc box
with $512^3$ dark matter particles  (solid line)  and the Sheth-Tormen 
approximation (dotted line)}  
\end{figure}

We proceed in the same manner to study the distribution of gas
particles. In particular, we run the MST 
procedure over all gas particles to obtain
catalogs of gas objects. Then we identify how gas and DM clusters are
related. In most cases this is straightforward, i.e. the centers
coincide.  In a few cases ($\sim 2$\%) centers do not coincide. The
simple reason is that the friends-of-friends algorithm may connect
the more smoothly distributed gas particles
by bridges which is not necessarily the case for the corresponding
DM particles. 

For our present analysis we define two cluster samples based on two
mass thresholds: (a) $M_{\rm cl}\geq 1.34 \times 10^{14}h^{-1}M_{\odot}$
(hereafter $\Lambda$CDM$_{1}$ sample) and (b) $M_{\rm cl}\geq 6.2 \times 
10^{13}h^{-1}M_{\odot}$ (hereafter $\Lambda$CDM$_{2}$ sample).  These
two subsamples contains 2773 and 7869 cluster entries with corresponding mean
densities of $n_{1} \simeq 2.1 \times 10^{-5}h^{3}\,$Mpc$^{-3}$ and
$n_{2}\simeq 6.1 \times 10^{-5}h^{3}\,$Mpc$^{-3}$, respectively.

\begin{table*}
\caption[]{List of the major parameters in the two analysed
  samples. The first column gives the model, column 2 gives the mass
  threshold, columns 3 and 4 give the clustering properties, column 5
gives the number of the simulated clusters 
contained in the volume used. Column 6 gives the number of
the superclusters with 10 or more members ($N_{\rm sup}^{10}$) and finally, column 7 gives
  the percolation radius used. 
Note that $r_{\circ}$, $R_{\rm pr}$ have
  units of $h^{-1}$Mpc and $M_{\rm lim}$ has units of $h^{-1}M_{\odot}$.}
\tabcolsep 12pt
\begin{tabular}{cccccccc} 
\hline  
sample & $M_{\rm lim}/h^{-1}M_{\odot}$ & $r_{\circ}$& $\gamma$ & $N_{\rm cl}$ & $N_{\rm sup}^{10}$&$R_{\rm pr}$\\ \hline \hline

$\Lambda$CDM$_{1}$    &  1.34$\times 10^{14}$  &  15.3$\pm 1.5$& 1.82$\pm 0.07$ &  2773 & 22& 18.0 \\

$\Lambda$CDM$_{2}$ &  6.2$\times 10^{13}$& 11.8$\pm 1.0$& 1.78$\pm0.04$ &  7869  & 54 & 11.5 \\ 
\hline     
\end{tabular}
\end{table*}

\begin{figure}
\label{fig:1}
\mbox{\epsfxsize=8.5cm \epsffile{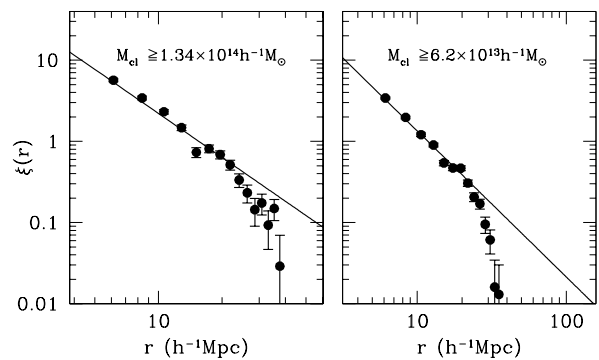}}
\caption{The spatial two-point correlation function 
of the $\Lambda$CDM$_{1}$ ($M_{\rm lim}=1.34\times 10^{14}h^{-1}M_{\odot}$) 
and 
$\Lambda$CDM$_{2}$ ($M_{\rm lim}=6.2\times 10^{13}h^{-1}M_{\odot}$) 
samples. 
The error bars are estimated using the bootstrap procedure. 
The dashed lines
represent the best-fitting power low $\xi(r)=(r_{\circ}/r)^{\gamma}$ 
(see parameters in Table 1).} 
\end{figure}
These mass thresholds were chosen in order to roughly 
reproduce the spatial density of the REFLEX
(B\"{o}hringer et al. 2001) and APM cluster samples (Dalton et al. 1994). 
Utilizing the classical correlation estimator described by Efstathiou et
al. (1991), we evaluate the real-space two-point correlation function,
$\xi(r)$, in logarithmic intervals. The resulting correlation function
parameters, for both cluster samples, are presented in Table 1. 
The derived correlation function
slope is very near to its nominal value of $\gamma=1.8$ and we find
$r_{\circ}=15.3\pm 1.4h^{-1}$Mpc and  
$r_{\circ}=11.8\pm 1.0 h^{-1}$Mpc, respectively.
It is clear that the correlation length increases with cluster richness, 
as expected from the well-known richness dependence of the correlation
strength (eg. Bahcall \& Burgett 1986; Bahcall \& West 1992).
In Fig. 2 we plot the derived two-point correlation
function as filled dots, while the best-fitting power law model 
$\xi(r)=(r_{\circ}/r)^{\gamma}$ is shown as straight lines (see Table 1).

The $\Lambda$CDM$_{1}$ sample has a correlation
length that within $\sim 1.5 \sigma$ resembles that of the REFLEX X-ray
cluster sample (Collins et al. 2000) while the $\Lambda$CDM$_{2}$ 
sample has a correlation length which approaches that of the APM
clusters (Dalton et al. 1994) and of a poor subsample of the SDSS-CE
clusters (Goto et al. 2002) analysed by Basilakos \& Plionis (2004).

\section{Structure Shape Determination}

\subsection{Defining $\Lambda$CDM Superclusters}
The identification of superclusters is based 
on the use of the friend of friends algorithm (otherwise called 
percolation algorithm; see Zeldovich et al. 1982), applied on the
distribution of simulated clusters. 
The algorithm starts by placing a sphere, of a certain radius, around each
cluster and then connects all neighboring spheres having an overlap region.
Doing so for all clusters in the sample the algorithm 
provides a unique, for each specific
percolation radius, list of connected clusters, dubbed
``superclusters''.
It is evident that different percolation radii will result in
different catalogues of superclusters. Therefore the choice of 
an optimal percolation radius is essential for the detection of 
superclusters which are related uniquely to the specific underline 
point (cluster) process. 

To this end, we use a ``critical'' value of the percolation
radius ($R_{\rm pr}$), which is related uniquely to the clustering 
properties of the underline cluster distribution, given by Peebles (2001): 
\be\label{eq:pee1}
R_{\rm pr} \simeq \left[\frac{3-\gamma}{\omega_{\rm s} n 
r_{\circ}^{\gamma}} \right]^{\frac{1}{3-\gamma}}\,\,,
\ee
where $\omega_{\rm s}$ is the solid angle covered by the cluster
sample, $n$ is the mean cluster
number density and $r_{\circ}$, $\gamma$ are the clustering parameters
of the cluster sample (see Table 1).

Thus, for the $\Lambda$CDM$_{1}$ and $\Lambda$CDM$_{2}$ cluster
samples, the critical percolation radius is $R_{\rm pr} \simeq 
18$ and $11.5\; h^{-1}$ Mpc, respectively.
Using these values, we apply the percolation algorithm to the whole simulation
box distribution of clusters, while in order to minimize edge effects,
related to the simulation box boundaries, we rap-around whenever a
supercluster touches the simulation borders.
The resulting number of superclusters, with 10 or more cluster members,
is 22 and 54, respectively for the $\Lambda$CDM$_{1}$ and
$\Lambda$CDM$_{2}$ cluster samples. In Fig. 3 we present a three
dimensional view of the detected ($\Lambda$CDM$_1$) superclusters
after smoothing them with a Gaussian window of a 6 $h^{-1}$ Mpc
radius.

\begin{figure}
\mbox{\epsfxsize=9.cm  \epsffile{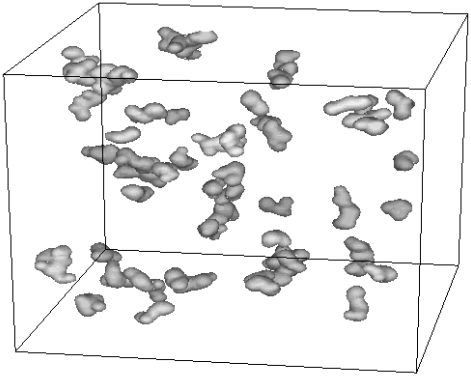}}
\caption{The 3D distribution of our detected superclusters (based on
  the $\Lambda$CDM$_1$ cluster sample). A Gaussian smoothing has been
  applied with 1$\sigma$ radius of 6 $h^{-1}$ Mpc. The elongated
  prolate-like nature of the detected superclusters is evident.}
\end{figure}

We have performed a further test in order to verify whether the
detected superclusters are the optimal ones. To this end we compare
the number of detected $\Lambda$CDM superclusters for different values
of the percolation radius with those detected, for the same radius, 
in random samples of clusters having the same space number
density as the simulation clusters. 
We run a large number of Monte-Carlo simulations in 
which we destroy the intrinsic $\Lambda$CDM clustering 
by randomizing the spatial coordinates of the clusters.
On this intrinsically random cluster distribution, 
we apply the procedure described before and identify the
random superclusters, $N_{\rm ran}$, which are due to our 
supercluster-identification method itself.

We define the probability that the detected supercluster in the
simulated data is real as ${\cal P} = 1-N_{\rm ran}/N_{\rm sup}$.  In
Fig. 4 we plot that probability depending on the percolation
radius for the $\Lambda$CDM$_{1}$ (solid line) and $\Lambda$CDM$_{2}$
(dashed line) cluster samples. If the percolation radius is
significantly smaller than the mean random intercluster separation then
the number of random supercluster is null and thus ${\cal P} \simeq 1$.
As the percolation radius increases then the number of random
superclusters increases as well, but with a slower rate than the
corresponding number in a clustered distribution.  However, there is a
trade-off which has to do with the fact that at extremely small
percolation radii, although we ensure that ${\cal P} \simeq 1$, we get
a very small number of superclusters. Therefore, the necessity of
having adequate statistics leads to a larger percolation radius.
\begin{figure}
\label{fig:2}
\mbox{\epsfxsize=9cm \epsffile{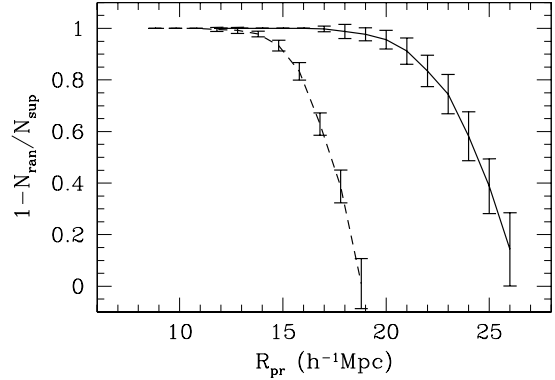}}
\caption{  
  The probability that the supercluster detected in the simulated data
  with 10 or more members is real as a function of the 
percolation radius $R_{\rm pr}$. The solid line is for the $\Lambda$CDM$_{1}$
  sample, the dashed for $\Lambda$CDM$_{2}$. }
\end{figure}
Since in the present study we wish to investigate the morphological
characteristics of superclusters and possible related environmental
effects, we limit our analysis to superclusters with more than 9
cluster members, for which we can robustly determine their shapes
(eg. Kolokotronis et al. 2002). 
We then find, for the percolation radii determined by eq.(\ref{eq:pee1}), ie.,
$R_{\rm pr}\simeq 18$ and $11\;h^{-1}$ Mpc for the $\Lambda$CDM$_{1}$
and the $\Lambda$CDM$_{2}$ samples respectively, that the significance
(which can be read from Fig. 4) is ${\cal P} >0.95$. 

An interesting property of our detected $\Lambda$CDM superclusters is
a correlation between the supercluster mass, defined as the sum of the
cluster members mass, and the cluster velocity dispersion within their
parent superclusters, of the form $M_{\rm sup}\propto
\sigma^{\alpha}$ (see Fig. 5). For superclusters with 4 or more
members we find that the linear
correlation coefficient in log-log space is $\sim 0.4$ with the
probability of random correlation being extremely small ($< 10^{-7}$)
for both $\Lambda$CDM$_1$ and $\Lambda$CDM$_2$ cluster samples.
The corresponding slopes of the $M-\sigma$ relation is $\sim 0.43\pm 0.08$ and
$\sim 0.56\pm 0.06$, respectively. Although the supercluster crossing
time along their major axis is much larger than the age of the
universe, $t_{u}$,
and thus superclusters have not yet turned around 
(see also Gramann \& Suhhonenko 2002), such a correlation imply that
they could be ``locally'' bound (eg. Barmby \&
Huchra 1998; Small et al. 1998). In fact, many superclusters have
crossing times along their minor axis less than $t_{u}$. This is shown
in Fig. 6
for the $\Lambda$CDM$_1$ (left panel) and $\Lambda$CDM$_2$ (right
panel) cluster samples.

Furthermore, it is interesting that supercluster scale peculiar velocities 
can produce distinct features on the CMB and thus provide the means 
for detection of superclusters at cosmological distances (eg.  Diaferio, Sunyaev \&
Nusser 2000).
\begin{figure}
\label{fig:M-s}
\mbox{\epsfxsize=8.5cm \epsffile{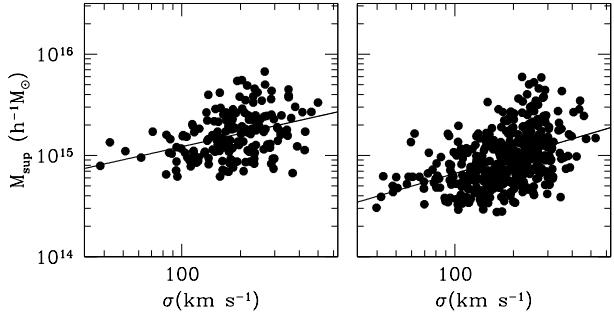}}
\caption{ The $M_{\rm sup}-\sigma$ relation for $\Lambda$CDM rich superclusters
(with 10 or more members). 
{\sc Left Panel.} Based on rich clusters ($\Lambda$CDM$_1$ sample).
{\sc Right Panel.} Based on poorer clusters ($\Lambda$CDM$_2$ sample).}
\end{figure}

\begin{figure}
\label{fig:Ms}
\mbox{\epsfxsize=8.5cm \epsffile{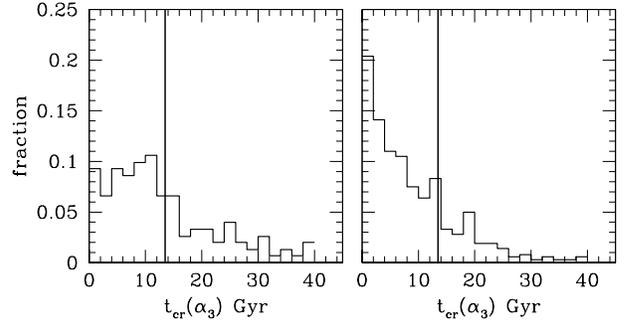}}
\caption{ The frequency distribution of superclusters with 4 or more
  members having their crossing times (along
  their minor axis) indicated along the x-axis.
{\sc Left Panel.} Based on rich clusters ($\Lambda$CDM$_1$ sample).
{\sc Right Panel.} Based on poorer clusters ($\Lambda$CDM$_2$ sample).}
\end{figure}

\subsection{Application of the Shapefinder Statistic}
To identify the characteristic morphological features of the large
scale structures, we will use a differential geometry approach, 
called ``shapefinders'' and introduced by Sahni et al. (1998).

The supercluster shapes are defined by fitting ellipsoids to the
distribution of cluster members.  In Kolokotronis et al. (2002) we
investigated the minimum number of points that are necessary in order
to determine unambiguously the true shape of a structure traced by a
point process. Here we estimate shapes for those superclusters that
consist of at least 10 cluster members.

In the case of clusters we can define the shape directly from the distribution
of dark matter or gas particles. In both cases we use the moments of inertia
method (eg. Carter \& Metcalfe 1980; Plionis, Barrow \& Frenk 1991).
Summing up over the DM or gas particles (for the case of clusters) 
or the cluster members (for the case of supercluster) we define 
the inertia tensor:

\begin{equation}\label{eq:diag}
 I_{\alpha\beta} = \sum_i m^{(i)} x^{(i)}_\alpha x^{(i)}_\beta.
\end{equation}
which after diagonalisation provides the eigenvalues $\alpha_1$,
$\alpha_2$, and $\alpha_3$ and the
corresponding eigenvectors which point into the direction of the
principal axes.
Since the eigenvalues can be considered as the three principal axes
of the triaxial ellipsoid with the volume $V=\frac{4\pi}{3} \alpha_1
\alpha_2 \alpha_3$ we can now relate the shape of the configuration 
to the shape of that ellipsoid.

Below we present the main features of the shape statistic, following the
notations of Sahni et al. (1998) (for a first application to
astronomical data see Basilakos et al. 2001).  To
characterize the shape of any object these authors introduced a set of
three quantities, defined by ${\cal H}_{1}=V S^{-1}$, ${\cal H}_{2}=S
C^{-1}$ and ${\cal H}_{3}=C$, where $V $ is the volume, $S$ the surface
and $C$ the integrated mean curvature of the given object. These
quantities have dimensions of length and are normalized to give ${\cal
  H}_{i}=R$ for a sphere of radius $R$.  With those quantities one
can define two dimensionless shapefinders: $K=(K_{1},K_{2})$,
where \be K_{1}=\frac{ {\cal H}_{2}-{\cal H}_{1} }{ {\cal H}_{2}+{\cal
    H}_{1} } \ee and \be K_{2}=\frac{ {\cal H}_{3}-{\cal H}_{2} } {
  {\cal H}_{3}+{\cal H}_{2} } \;\;.  \ee

The above technique characterizes the shapes of topologically 
non-trivial cosmic objects according to the following classification:
\begin{enumerate}
\item Pancakes if $K_{1}/K_{2}>1$
\item Filaments if $K_{1}/K_{2}<1$
\item Ribbons for $(K_{1},K_{2})\simeq (\alpha,\alpha)$
with $\alpha \le 1$ and thus
structures $K_{1}/K_{2}\simeq 1$ and
\item Spheres if $\alpha_{1}=\alpha_{2}=\alpha_{3}$ and thus $K_{1}=
  K_{2}=0$
\item ideal filament (one-dimensional objects) for 
${\cal H}_{1} \simeq {\cal H}_{2} <<{\cal H}_{3}$ 
\item ideal pancake (two-dimensional objects) for 
${\cal H}_{1} << {\cal H}_{2} \simeq {\cal H}_{3}$ 
\item ideal ribbon for 
${\cal H}_{1} << {\cal H}_{2} << {\cal H}_{3}$ 
\end{enumerate}
Ideal filaments (0,1), pancakes (1,0), ribbon structures (1,1)
and spheres (0,0) are represented by the four vertices of the shape plane in
the form of the shape vector $K=(K_{1}$, $K_{2}$), whose amplitude and
direction determines the morphology of an arbitrary 3D surface. 
Finally, for the quasi-spherical objects the $K_{1,2}$ are small and
therefore their ratio ($K_{1}/K_{2}$) measures the deviation from 
perfect spherical shapes. Note that below we will consider as
spherical all simulated objects that have $K_1, K_2 <0.005$, the axis
ratios of which lie in the range: 
$1\le \alpha_{1}/\alpha_{3}\le 1.20$ with 
$\langle \alpha_{1}/\alpha_{3}\rangle \simeq 1.10 \pm 0.06$ 
and
$1\le \alpha_{1}/\alpha_{2}\le 1.25$ with 
$\langle \alpha_{1}/\alpha_{3}\rangle \simeq 1.15 \pm 0.05$. 

\subsubsection{The Shape of DM and Gas Cluster Halos}
\label{resol}
Using the above procedure we have estimated the shape of both the Dark
Matter and the gas component of the $\Lambda$CDM clusters. We have
performed this analysis for two minimal spanning tree linking
parameters: for the nominal value of 0.17 and for $0.17/2^n$ with
$n=1$ (probing densities $8^n$ times larger than in the former case
and hence the inner dense cluster cores).
In Fig. 7 we present the corresponding shape spectrum for the
$\Lambda$CDM$_1$ (left panel) and the poorer $\Lambda$CDM$_2$ cluster
samples (right panel). The main panels show the results based on the
Dark Matter cluster component while the inner panels to the
corresponding gas component. 
The DM cluster component of both the $\Lambda$CDM$_1$ and
$\Lambda$CDM$_2$ cluster samples show a
strong preference for filamentary-like (prolate-like) configurations
with $\sim$ 75\% and 72\% of the clusters having $K_1/K_2 <1$ while
only 1.7\% and 2.5\% can be considered spherical.
Note that in the main panels of Fig.7 the 
shaded histogram corresponds to the larger
linking length while the dashed line to the smaller one. The results
based on the Dark Matter cluster component are very similar for both
linking parameters indicating that both the inner cluster dark matter core
and the cluster outskirts have a consistent shape.
\begin{figure}
\mbox{\epsfxsize=8.cm \epsffile{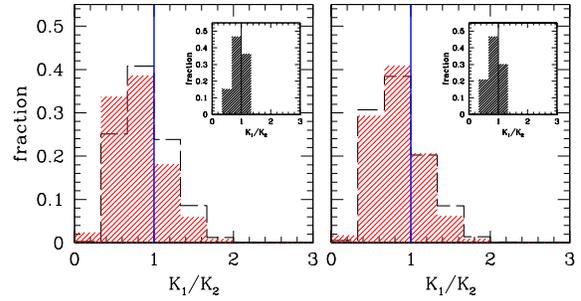}}
\caption{The cluster halo shape spectrum: {\sc Left Panel.} Rich
  clusters ($\Lambda$CDM$_1$ sample), {\sc Right Panel.} Poorer
  clusters ($\Lambda$CDM$_2$ sample). The shaded histograms correspond
  to a linking parameter of 0.17 ($\rho/\rho_{\rm mean}\simeq 330$)
  while dashed lines to the denser (by a factor of $\sim 8$) 
  core cluster region (linking parameter of 0.085). The insert 
  shows the corresponding results for the gas cluster component.}
\end{figure}

Regarding the gas cluster component the results are quite
different. As can bee seen from the insert of Fig.7 the shape
spectrum peaks at $K_1/K_2\simeq 1$ with little dispersion around this
value. This result together with the quite small values of the $K_1$
and $K_2$ parameters indicate that the gas distribution is much more
spherical than the corresponding dark matter component, as expected.
We find that the fraction of clusters having a spherical gas component
is 53\% and 64\% for the $\Lambda$CDM$_1$ and
$\Lambda$CDM$_2$ cluster samples respectively, while for the smaller
linking parameter (the dense cluster cores) the corresponding values
are $\sim 44\%$ and $49\%$. This indicates that the gas distribution
in the cluster cores is more flattened than in the cluster outskirts.
Furthermore, the sphericity of the gas halos implies that the position 
angles orientations of
gas halos are ill-defined and thus should be used in alignment studies
with great caution. To visualize this effect we present in Fig. 8
the misalignment angle between the major axes of the DM and gas mass
halos as a function of the ratio of the gas halo minor to major axis. 
Values of the
latter near unity imply spherical gas halo shapes. It is evident
that indeed the large misalignment angles are correlated with the gas
halo sphericity.
\begin{figure}
\mbox{\epsfxsize=8.5cm \epsffile{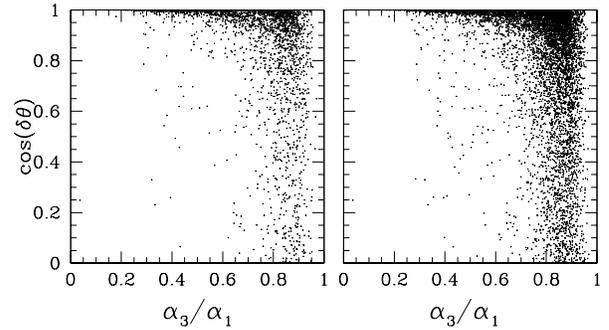}}
\caption{Gas halo sphericity versus the misalignment angle between
  the major axis of the DM and gas cluster halo components for the
  main linking parameter used (0.17). 
{\sc Left Panel.} Rich
  clusters ($\Lambda$CDM$_1$ sample), {\sc Right Panel.} Poorer
  clusters ($\Lambda$CDM$_2$ sample).}
\end{figure}

Finally, in order to test possible resolution effects on the halo
shapes, we present in Fig. 9 a comparison of the 
DM and gas halo axis ratio distribution ($M_{DM}>6\times
10^{13}$ $M_{\odot}$)
in  the 512$^3$ simulation  with the same quantities obtained from  another
run of the same simulation with 8 times coarser mass resolution  ($2\times
256^3$ gas and DM particles, see \S \ref{nbody} for details)   
 It is evident that the DM axis ratio distributions (left
panel) are indistinguishable while there is a small effect for the
case of the gas halos (right panel), in the direction of having 
artificially lower gas halo sphericity in the lower resolution simulation.
\begin{figure}
\mbox{\epsfxsize=8.5cm \epsffile{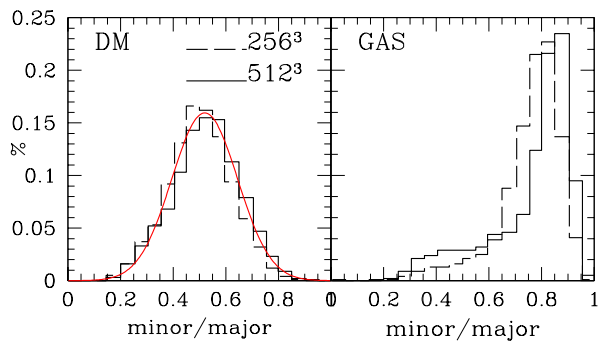}}
\caption{Comparison of the DM and Gas halo sphericity as a function of
  simulation resolution. In the left panel we also plot, as a
  continuous line, the Jing \&  Suto (2002) model which fits well the
  DM halo axis ratio distribution.}
\end{figure}

\subsubsection{The Shape of Superclusters}
In Fig.10 we show the shape spectrum of the detected
superclusters (with 10 or more members -
dashed-line histograms) and we compare with the corresponding spectrum of
their cluster members (for a linking parameter of 0.17).
It is obvious that the dominant shape of cosmic
structures, being clusters or superclusters, is that of prolate
structures (filaments), in agreement with previous theoretical and
observational studies (eg. Plionis et al. 1991; Basilakos, Plionis \&
Maddox 2000; Sathyaprakash et al. 1998; Basilakos et al. 2001;
Kolokotronis et al.  2002; Basilakos 2003; Einasto et al. 2003).
Furthermore, it is clear that superclusters are more flattened
structures than clusters themselves, populating a larger region in the
shape-spectrum.

We have verified the dominance of filamentary supercluster shapes for
the ``optimal'' percolation radius, given by eq. (\ref{eq:pee1}),
although it appears that superclusters have a slightly lower 
filamentariness with respect to their cluster members. The relevant
fractions are: filaments ($\sim 64\%$), pancakes ($\sim 32\%$)
and ribbons ($\sim 4\%$). 
Note that there are no spherical superclusters.
\begin{figure}
\mbox{\epsfxsize=8.5cm \epsffile{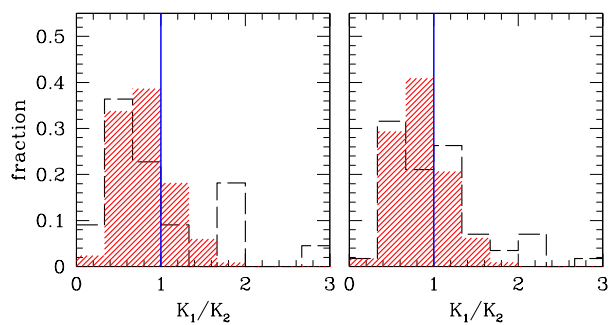}}
\caption{Comparison of the shape spectrum of superclusters containing
  10 or more clusters (dashed-line histogram) and of the DM clusters for the two
  subsamples (left and right figure). }
\end{figure}

We have further investigated the dependence of the supercluster shape
on the value of the percolation radius and we present our results in
Fig. 11 where we plot the fraction of superclusters having
a filament-like shape ($K_1/K_2<1$).
It is evident that for the richer cluster sample (left panel of
Fig. 11), the filamentariness
is a dominant feature only around the ``optimal'' percolation radius
(indicated by the arrow). Similarly, for the poorer cluster sample
(right panel) the ``optimal'' percolation radius coincides with a
local maxima in the filamentary fraction. However, in this cluster
sample, the filamentariness appears to be a generic feature, almost
independent of the percolation radius used.
\begin{figure}
\mbox{\epsfxsize=7.8cm \epsffile{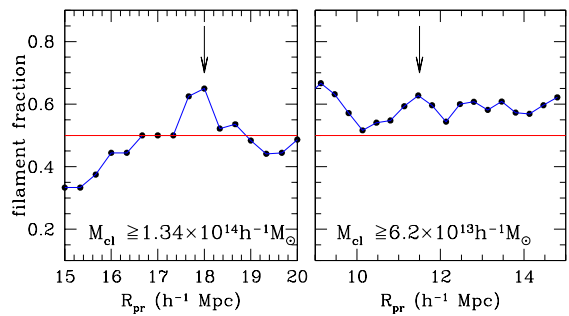}}
\caption{The fraction of filamentary superclusters as a function of
  percolation radius used to define the superclusters. The arrow
  indicates the ``optimal'' percolation radius given by eq.(\ref{eq:pee1}).
{\sc Left Panel.} Rich
  clusters ($\Lambda$CDM$_1$ sample), {\sc Right Panel.} Poorer
  clusters ($\Lambda$CDM$_2$ sample).}
\end{figure}

\section{The Environmental Trends of the Large Scale Network}
\subsection{Correlations of vector alignments}
As already mentioned in the introduction, an interesting question
regarding environmental effects on large scales is whether clusters are
aligned with the orientation of their parent supercluster.  To study
the possible supercluster - clusters alignment we attach to each cosmic
structure (being either supercluster or cluster) the orientation as
three unit vectors along the three main axes. 
\begin{table*}  
\caption[]{Correlation analysis results for the different pairs $x,y$
using the cluster DM component. The straight lines in Fig.12 represent 
the least-square fit to the data and have the general form $y=\lambda x+\beta$.
Also, ${\cal R}$ is the correlation coefficient with the
corresponding probability (${\cal P}_{R}$)
of random correlation. Note that $M_{\rm sup}$ has 
units of $10^{14} h^{-1}M_{\odot}$.}

\tabcolsep 12pt
\begin{tabular}{cccccc} 
\hline  

Sample & Pair $x,y$ & $\lambda$& $\beta$ & ${\cal R}$ & ${\cal P}_{R}$ \\ \hline \hline

$\Lambda$CDM$_{1}$ & $K_{1}/K_{2}$, ${\cal A}(\alpha_{1})$ & -0.04$\pm 0.02$  &  0.63$\pm 0.02$ & -0.39 & 7.9$\times 10^{-2}$ \\

 & $K_{1}/K_{2}$, $M_{\rm sup}$ & 9.3$\pm 4.2$  &  31.2$\pm 4.4$ & 0.45 & 3.8$\times 10^{-2}$ \\


$\Lambda$CDM$_{2}$ & $K_{1}/K_{2}$, ${\cal A}(\alpha_{1})$ & -0.05$\pm 0.02$  &  0.60$\pm 0.02$ & -0.29 & 3.6$\times 10^{-2}$ \\

  & $K_{1}/K_{2}$, $M_{\rm sup}$ & -  &  - & -0.11 & 0.40 \\
\hline     
\end{tabular}
\end{table*}

The distance vector between the supercluster center of mass and the
cluster member is ${\bf r}$, and the normalized direction is ${\hat{\bf
    r}}={\bf r}/r$. Based on the notations of Beisbart et al. (2002) we
consider the following alignment estimator (see also Stoyan \& Stoyan
1994; Faltenbacher et al. 2002): \be A(r)= \langle |{\bf e_l}
\cdot {\bf e_k}| \rangle (r) \;\;.  \ee Therefore, for each
supercluster $l$, the factor $A(r)$ describes the direct alignment of
the corresponding vector ${\bf e_l}$ with the vectors ${\bf e_k}$ of
the clusters within the parent supercluster.  Note that $A(r)$ is
proportional to the cosine of the angle between ${\bf e_l}$ and ${\bf
  e_k}$. The case of a random alignment signal implies $A(r)=0.5$.  


\subsection{Shape-Alignment Correlation}
In Fig.12 (top panels) we present the alignment $A(\alpha_{1})$ 
estimator as a function of the supercluster shape, given by the ratio
$K_{1}/K_{2}$. 
There is an obvious supercluster shape-alignment relation,
seen in both cluster catalogues ($\Lambda$CDM$_1$ and
$\Lambda$CDM$_2$), with supercluster alignment increasing
with supercluster filamentariness. Furthermore, only for the high
cluster mass sample, there is also a significant correlation between
the supercluster mass and shape, with mass decreasing with increasing
filamentariness. Probably, this is to be expected since 
pancake-like structures,
being two-dimensional should be more massive than the one-dimensional
filaments. However this is not observed in the case of $\Lambda$CDM$_2$.

\begin{figure}
\label{fig:5}
\mbox{\epsfxsize=8cm \epsffile{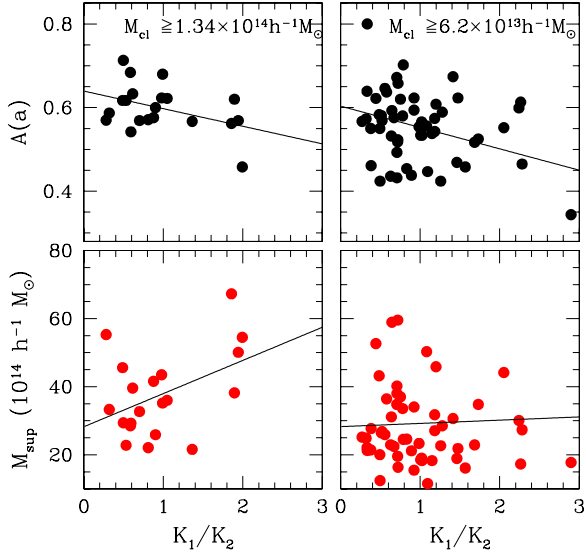}}
\caption{The shape-alignment (upper panels) and shape-mass (lower
  panels) correlations for the two supercluster samples analyzed and
  at $a=0.8 \alpha_1$.
The solid lines are the best least-square fits to
the data (see Table 2).}  
\end{figure}
Table 2 summarizes the quantitative correlation results for all tests
using the cluster DM halo orientations.
Probably due to the small number of superclusters
(especially in the higher cluster mass sample) the
significance of the correlations is not very high. In order to test
whether the signal could be affected by a few cluster outliers at the
supercluster edges, we have derived  the correlation signal and its
significance as a
function of the parametrized distance of the member clusters from the
supercluster center of mass (ie., as a function of $r/\alpha_1$, where
$r$ is the physical size of the supercluster and $\alpha_1$ is its
major axis). In Fig.13 we present our results from which it is
evident that the signal is relatively stable, although it appears to
be higher and more significant around $r/\alpha_1\sim 0.8$, indicating that
indeed at larger distances, some cluster outliers do reduce the
correlation signal.

We have also used the gas halo orientations to repeat the same
analysis but as expected from the results presented in section 3.2.1
(see Fig. 8) we do not find a significant signal for the
$\Lambda$CDM$_1$ sample. However, for the poorer $\Lambda$CDM$_2$
sample we have found a correlation signal (${\cal R}=-0.25$) with a
random probability of ${\cal P}\simeq 0.07$.
\begin{figure}
\label{fig:7}
\mbox{\epsfxsize=8.5cm \epsffile{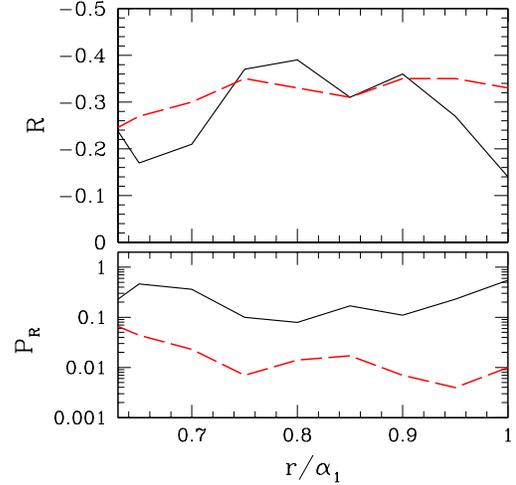}}
\caption{The shape-alignment correlation coefficient (upper panel) and
  random probability (lower panel) as a function of
  supercluster-centric distance.} 
\end{figure}

Furthermore, we have seen in Fig. 11 that the fraction of filamentary
superclusters depend on the percolation radius used and thus if, as
suggested by the correlations in Table 2 and Fig. 12, the alignment
signal is higher in the filamentary superclusters we may expect a drop of
the signal when the distribution of superclusters is dominated by
pancake-like morphologies.
Therefore, in order to investigate the above as well as the 
robustness of the correlation signal we re-assigned
cluster DM halos to superclusters using a range of percolation radii, around
the ``optimal'' value given by eq.(\ref{eq:pee1}). In Fig. 14 we
present both the value of the shape-alignment correlation coefficient
and its random probability as a function of percolation radius used to
define the superclusters. The resulting correlation coefficients is
higher and more significant very near the ``optimal'' percolation
radius, defined by eq.(\ref{eq:pee1}), which also coincides with peaks in the
distribution of supercluster filamentary fraction (Fig.11).

\begin{figure}
\label{more}
\mbox{\epsfxsize=8.5cm \epsffile{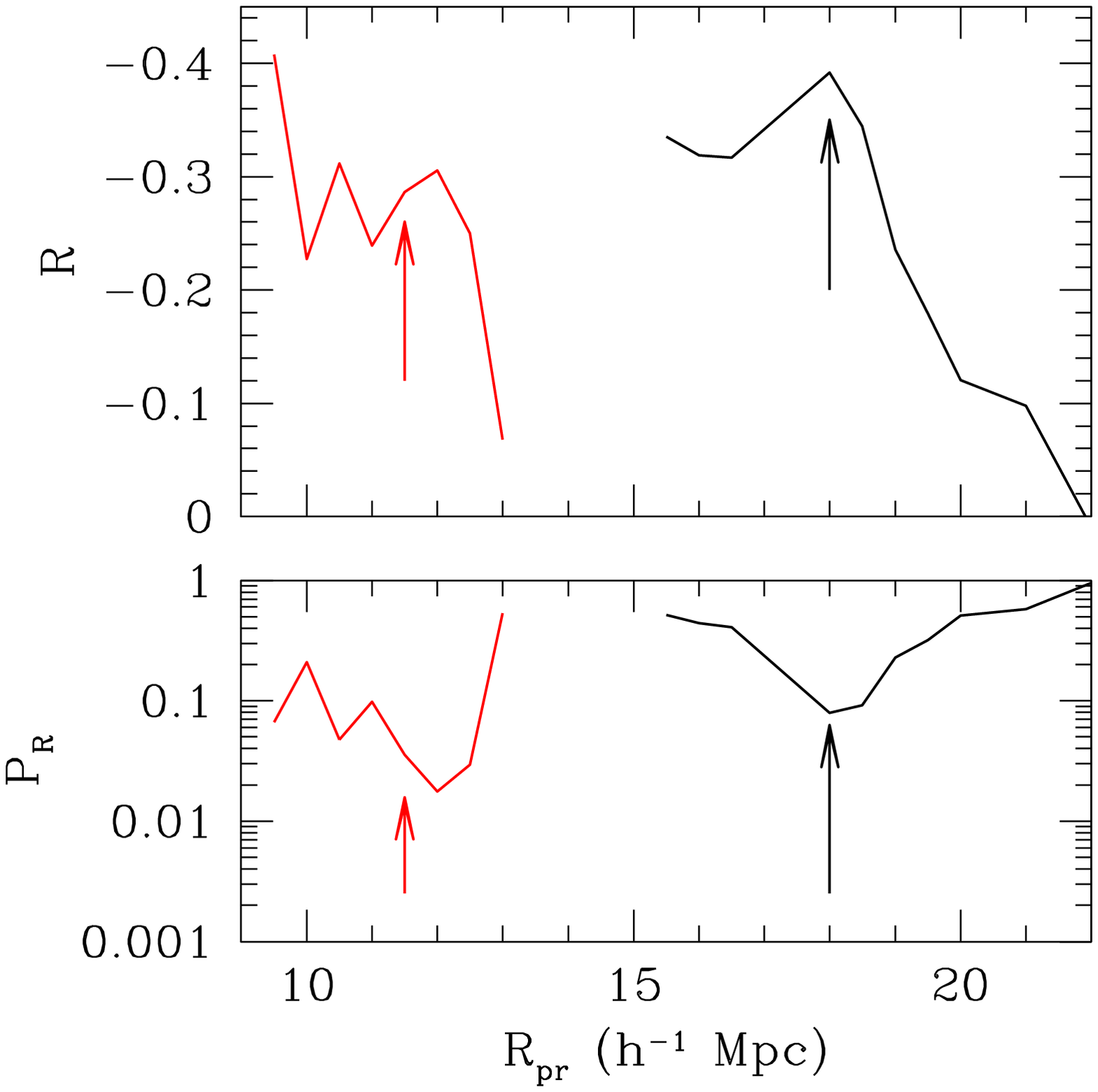}}
\caption{The shape-alignment correlation coefficient (upper panel) and
  random probability (lower panel) as a function of percolation radius
  used to define the superclusters. The arrow indicates the
values corresponding to the  ``optimal'' percolation radius. The
  graphs on the right and left correspond to the lower ($\Lambda$CDM$_2$)
  and higher ($\Lambda$CDM$_1$) mass samples.} 
\end{figure}
The increase of the supercluster filamentariness with alignment, 
is to be expected according to the hierarchical clustering scenario,
where gas and galaxies flow into denser regions along anisotropic
directions, defined by the large-scale structure
(cf. West 1994; Ostriker \& Cen 1996 and references therein). As a result of
these inflows we expect clusters to be aligned with their neighbours, 
especially when both are members of the same filament. 
For low-$\Omega_{\rm m}$ cosmologies
one expects that major merging and anisotropic accretion of matter along
filaments will have stopped long ago. Thus
gravitational violent relaxation would tend to isotropize the cluster 
phase-space configuration, more so in the recent times. However, it
has been shown that clusters retain memory of their initial anisotropic configuration
from were they accreted the main bulk of the matter that they contain
(eg. van Haarlem \& van de Weygaert 1993).

\subsection{Supercluster - cluster velocity correlations}
In this section we investigate possible correlations 
between the supercluster shapes and the alignment between
supercluster major axis and their member cluster velocity field. 
Using the notations of section 4.1 
we can define the following cross-correlation vector estimator:
 \be
U(r)= \langle |{\bf e_l} \cdot {\bf v_k}|  \rangle (r) 
\ee 
where $\bf v_{k}$ is velocity vector with 
$|{\bf v_{k}}|=1$, which corresponds to each host cluster while 
${\bf e_{l}}$ is the eigenvector of the major ($r=\alpha_{1}$)
supercluster axis.

We find that shape - velocity alignment correlations does not exist at
any significant level. However, for the case of high mass clusters
($\Lambda$CDM$_1$ sample) there is a relatively significant
correlation between $A(r)$ and $U(r)$, indicating that the alignment
between supercluster and cluster member major axes is accompanied with a
similar alignment between supercluster major axis and the direction of
the cluster member peculiar velocity (see Fig. 15).


\begin{figure}
\mbox{\epsfxsize=10.5cm \epsffile{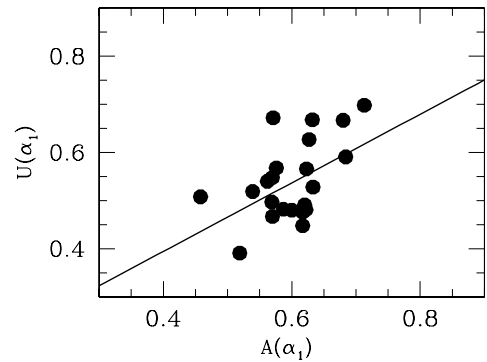}}
\caption{The cluster velocity -supercluster alignment $[U(r)]$ 
and cluster major axis-supercluster alignment $[A(r)]$ correlation for
the $\Lambda$CDM$_1$ superclusters.}
\end{figure}

\section{Conclusions}
We have studied the morphological features of $\Lambda$CDM cluster 
dark matter and baryonic gas halos  as well as superclusters 
using a large volume ($V=500^{3} \;h^{-3}$ Mpc$^{3}$) N-body+GADGET 
simulation with unprecedented mass resolution. The
measure of the structure geometry has shown that
prolate-like (filamentary) shapes dominates over pancakes, 
in agreement with other recent large-scale structure studies.
We have also presented evidence that there is a specific link between 
the orientation of cluster dark matter halos and that of the large scale
network. Cluster size halos appear to be aligned with their parent 
supercluster major axis, more so if the supercluster is
filamentary-like. 
For the richest cluster halos we have also found a correlation between
the halo peculiar velocity -
supercluster alignment and the cluster major axis-supercluster
alignment signals. 

\section* {Acknowledgments}
We thank the anonymous referee for his/her critical comments and useful 
suggestions.
SG thanks DAAD for supporting our collaboration. VT
thanks DFG for supporting his visit at AIP. Simulations were done 
at the John von Neumann Institute for Computing J\"ulich (Germany).
GY and MP thanks the {\em Plan Nacional de Astronom\'\i a y Astrof\'\i sica}
of Spain for financial support under project number AYA2003-07468.
Furthermore, MP acknowledges support by the Mexican Government grant No
CONACyT-2002-C01-39679. 

\end{document}